\documentclass[12pt,preprint]{aastex}
\usepackage{amsmath,euscript,amsfonts}

\newcommand{\beq}{\begin{equation}}
\newcommand{\eeq}{\end{equation}}
\newcommand{\ba}{\begin{array}}
\newcommand{\ea}{\end{array}}

\begin{document}


\lefthead{Liu et al.} \righthead{Model considerations and
calculations for the multi-wavelength emission of transient Swift
J164449.3+573451}

\title{
A two-component jet model for the tidal disruption event Swift
J164449.3+573451}

\author{\small \it Dangbo Liu\altaffilmark{1,2}, Asaf Pe'er\altaffilmark{1,3}, Abraham Loeb\altaffilmark{1}
%
\affil{$^{1}$Institute for Theory and Computation, Harvard-Smithsonian
  Center for Astrophysics, 60 Garden Street, Cambridge, MA 02138, USA}
\affil{$^{2}$Center for Astronomy and Astrophysics,
  Department of Physics and Astronomy and Shanghai Key Lab for Particle Physics and
  Cosmology, Shanghai Jiao Tong University, 800 Dongchuan Road,
  Shanghai, 200240, China}
\affil{$^{3}$Physics Department, University College Cork, Cork,
  Ireland} }

\begin{abstract}
We analyze both the early and late time radio and X-ray data of the
tidal disruption event Swift J1644+57. The data at early times
($\lesssim 5$~days) necessitates separation of the radio and X-ray
emission regions, either spatially or in velocity space. This leads
us to suggest a two component jet model, in which the inner jet is
initially relativistic with Lorentz factor $\Gamma \approx 15$,
while the outer jet is trans-relativistic, with $\Gamma \lesssim
1.2$. This model enables a self-consistent interpretation of the
late time radio data, both in terms of peak frequency and flux. We
solve the dynamics, radiative cooling and expected radiation from
both jet components. We show that while during the first month
synchrotron emission from the outer jet dominates the radio
emission, at later times radiation from ambient gas collected by the
inner jet dominates. This provides a natural explanation to the
observed re-brightening, without the need for late time inner engine
activity. After $100$~days, the radio emission peak is in the
optically thick regime, leading to a decay of both the flux and peak
frequency at later times. Our model's predictions for the evolution
of radio emission in jetted tidal disruption events can be tested by
future observations.
\end{abstract}

\keywords{black hole physics --- galaxies: jets --- galaxies: nuclei
--- radiation mechanisms:non-thermal}


\section{Introduction}
\label{sec:intro}

A stray star, when passing near a massive black hole (MBH) can be
torn apart by gravitational forces, leading to a tidal disruption
event (TDE). Such an event would be observed as bright emission from
a previously dormant MBH, as it is being fed by temporary mass
accretion established after the tidal disruption of a passing star
\citep{Hills75, Rees88, EK89}. On 28 March 2011, an unusual
transient source Swift J164449.3+573451 (hereafter Sw J1644+57) was
reported, potentially representing such an event \citep{Burrows+11,
Levan+11}. This event was found to be in positional coincidence
($\lesssim 0.2$~kpc) with a previously dormant host galaxy nucleus,
at redshift $z = 0.354$ \citep{Levan+11, Fruchter+11, Berger+11}.

The rapid variability seen in the X-rays, of $\sim 78$~s
\citep{Burrows+11, Bloom+11}, implies a compact source size
$\lesssim 0.15$~AU, which is a few times the Schwarzschild radius of
$10^6 M_\odot$ MBH \citep[see also][]{MG11}. When combined with the
very high $\gamma$-ray and X-ray luminosity, $\approx 10^{47} \,
{\rm erg \, s^{-1}}$ \citep{Burrows+11, Bloom+11}, which is 2 - 3
orders of magnitude above the Eddington limit of such a MBH, it was
concluded that the X-ray emission must originate from a relativistic
jet of Lorentz factor $\Gamma \gtrsim 10$ \citep{Bloom+11}
\citep[see, however,][for
  alternative models]{KP11, OSJ11, QK12, Socrates12}.

While early works that investigated the expected signal (optical/UV
emission) from such an event were focused on the signal from the
accreting material of the stellar debris \citep{Rees88, LU97, Ulmer99,
  Bogdanovic+04, GRRK09, SQ09}, recently, the observational signature
from a newly formed jet was considered \citep{GM11, VKF11, WC12,
  MGM12, Decolle+12, SL12}. The basic mechanism suggested in these
works is similar to the mechanism that is thought to operate in
gamma-ray bursts (GRBs), namely energy dissipation by either internal
shock waves \citep{RM94, PX94} or by an external (forward) shock wave,
accompanied at early stages by a reverse shock wave \citep{RM92, MR93,
  PSN93, SP95}. These shock waves, in turn, are believed to accelerate
particles and generate strong magnetic fields, thereby producing
synchrotron radiation, accompanied by synchrotron self Compton (SSC)
emission at high energies.

Indeed, shortly after its discovery, an extensive radio campaign
showed that the X-ray emission is accompanied by bright radio
emission \citep{Zauderer+11}, which was interpreted as synchrotron
emission from the jetted material, thereby supporting this
hypothesis. However, a careful analysis revealed that the radio
emitting material propagates at a more modest Lorentz factor,
$\Gamma \gtrsim 1$ \citep{Zauderer+11}, and therefore the X-ray and
radio emission cannot have similar origin. This had led to the
suggestion that the X-rays may originate from internal dissipation
\citep{WC12}, while the radio may originate from the forward shock
that propagates into the surrounding material \citep{MGM12, CW12}.

Late time radio monitoring, extending up to $\approx 582$ days
\citep{Berger+12, Zauderer+13}, revealed an unexpected behavior:
after about $\sim 30$~days (observed time), the radio emission
showed re-brightening, which lasted up to $\sim 100$~days, after
which the radio flux decayed. This re-brightening was not
accompanied by re-brightening in the X-ray flux, and is not expected
in the context of the forward shock models. This led
\citet{Berger+12} to conclude that the radio re-brightening resulted
from late time energy injection (however, for alternative views, see
\citet{Kumar+13, BP13}). Thus, a comprehensive model that considers
the temporal as well as combined radio and X-ray spectral data is
still lacking.

Any such model must take into account the fact that the emitting
material is mildly relativistic at most. First, the radio emitting
material propagates at Lorentz factor $\Gamma \gtrsim 1$. Second,
while the X-ray emitting material propagates at initial Lorentz factor
$\Gamma \simeq 10 - 20$, its velocity becomes trans-relativistic on
the relevant time scale of tens of days, as surrounding material is
collected. The dynamics of such trans-relativistic propagation was
recently considered by \citet{Peer12}.

Here, we propose a new model that considers simultaneously the
emission of both the radio and the X-rays, their spectrum as well as
temporal evolution. We re-derive the constraints set by both radio
and X-ray observations, and confirm that indeed at early times
(first few days) these must have a separate origin.
%
%
We calculate the dynamics of the X-ray emitting plasma as it collects
material from the surrounding and decelerates.  We show that after
$\sim 30$~days (observed time), synchrotron emission from this plasma
peaks at radio frequencies, thereby providing a natural explanation to
the re-brightening seen at radio frequencies at these times, without
the need for late time internal engine activity. Moreover, we show
that the decay of the radio flux after $\sim 100$ days is naturally
explained by synchrotron self absorption. At this stage, the flow is
in the trans-relativistic regime ($\Gamma - 1 \simeq 1$).


This paper is organized as follows. In \S\ref{sec:early_times}, we
carefully revise the observed data of both the radio and X-ray
emission at early times (up to five days). While our treatment is
more general than previous works, we confirm earlier conclusions
that indeed the radio and X-ray emission must have separate origins.
In \S\ref{sec:late_times} we consider the temporal evolution of the
X-ray emitting material as it slows and cools, and show that it can
be the source of the re-brightening at radio frequencies seen after
$\sim 30$~days. We further consider radiative cooling in
\S\ref{sec:cooling}. We show that the very late ($\gtrsim 100$~days)
decay of the radio flux is naturally attributed to emission in the
optically thick regime: as the electrons cool, eventually the peak
of the synchrotron becomes obscured.  We compare our model to the
late time radio data of Sw J1644+57 using two scenarios: spherical
expansion and lateral expansion in \S\ref{sec:data}, before
summarizing our main results in \S\ref{sec:summary}.

%



\section{Early time radio and X-ray emission and its interpretation}
\label{sec:early_times}

Sw J1644+57 is a long-lived (duration $\gtrsim$ months) X-ray
outburst source, accompanied by bright radio emission, interpreted
as synchrotron radiation.  During the first few days of
observations, the isotropic X-ray luminosity ranged from $\sim
3\times 10^{45} \, {\rm
  erg \, s^{-1}}$ to a peak as high as $\sim 3 \times 10^{48} \, {\rm
  erg \, s^{-1}}$ \citep{Burrows+11} with average X-ray luminosity
$\approx few \times 10^{47}\, {\rm erg \, s^{-1}}$. The X-ray
emission peaked at frequency $\sim 2 \times 10^{18}$~Hz, with
estimated uncertainty up to two orders of magnitude above this value
\citep{Zauderer+11}. During the first few days, the X-ray lightcurve
was complex and highly variable, with variability time scale as
short as $\sim 100$~s \citep{Burrows+11, Levan+11}. On the high
frequency side ($\sim$ GeV emission), the Fermi LAT upper
limits are two orders of magnitude below the X-ray luminosity
\citep{Burrows+11}.

This source triggered a radio campaign, that began a few days after
its initial discovery. Radio observations showed that the peak
frequency occurs at $\sim 8\times 10^{11}\eta $~Hz with uncertainty
$1\leq \eta\leq 10$, and peak luminosity $\nu L_{\nu} \gtrsim
10^{43.5} \, {\rm erg \,
  s^{-1}}$ during the first few days. The spectral energy distribution
(SED) at the radio band ($< 345$~GHz$)$ at $\Delta t^{\rm ob} \approx
5$~days is well described by a power law, $F_{\nu}\propto\nu^{1.3}$ up
to $F_{\nu}(|_{\nu =345 {\rm GHz}}) \approx 35$~mJy. The steep power
law index requires self-absorbed synchrotron emission, with self
absorption frequency $\nu_{a}\gtrsim 10^{11}$~Hz \citep{Zauderer+11,
  Berger+12}. Within two weeks, the X-rays maintained a more steady
level, albeit with episodic brightening and fading spanning more
than an order of magnitude in flux, while the low frequency (radio)
emission decreased markedly in this period \citep{Levan+11,
Zauderer+11}.


In this section, we interpret the publicly available radio and X-ray
data at 5 days in the framework of the synchrotron (radio) and the
inverse Compton (X-rays) model. As we show below, in spite of the
relatively large number of free model parameters, the constraints
set by the existing early time data exclude this model: {\it the
same electrons cannot be responsible for simultaneous emission of
both the radio and X-ray photons in the framework of this model.} We
provide in table 1 below constraints on (some of the) free model
parameters derived from existing data, and show that either of the
three: the emission radius (corresponding to observed time of 5
days), the bulk motion Lorentz factor or the emitting electrons
characteristic Lorentz factor cannot be the same for the two (radio
and X-rays) emitting regions. This leads us to suggest an
alternative model, that of the structured jet, to be discussed
below.

\subsection{Interpretation of early radio emission}
\label{sec:radio1}

The radio emission observed from Swift J1644+57 is assumed to have
synchrotron origin. The synchrotron emitting plasma can be described
by five free parameters: the source size $R$, bulk Lorentz factor
$\Gamma$, total number of radiating electrons $N_{\rm e}$, magnetic
field strength $B$, and the characteristic electron Lorentz factor
(as is measured in the plasma frame), $\gamma_{e}$. Calculations of
the values of these parameters appear in \citet{Zauderer+11}. Here,
we generalize the treatment in \citet{Zauderer+11} by removing the
equipartition assumption used in that work.

Existing data provides the following four constraints.  The
observed characteristic frequency and total luminosity of synchrotron
emission, $\nu_{m}^{\rm ob}$ and $\nu L_{\nu}$ are given by \citep{RL79}
\begin{equation}
\ba{lcl}
\nu_{m}^{\rm ob} & = & \frac{3}{4\pi}\frac{q_{e}B}{m_{\rm e}c}\gamma_{e}^{2}\Gamma
=4.20\times 10^{6}\; B\, \gamma_{e}^{2}\, \Gamma  = 8 \times 10^{11}~{\rm Hz}; \\
\nu L_{\nu} & \approx & N_{e}P_{\rm
  syn}=\frac{4}{3}N_{e}\sigma_{T}c\gamma_{e}^{2}U_{B}\Gamma^{2} =
1.06\times 10^{-15}\; N_{e}\, \gamma_{e}^{2}\, B^{2}\, \Gamma^{2} = 3
\times 10^{43}~{\rm erg~s^{-1}}.
\ea
\label{totalflux}
\end{equation}
Here and below, $P_{\rm syn}$ is the total synchrotron radiation
power, $\sigma_{T}$ is the Thomson cross section, $U_{B}\equiv
B^{2}/8\pi$ is the magnetic energy density and CGS units are used.
In deriving equation (\ref{totalflux}), spherical explosion was
assumed.

A third condition is the synchrotron self absorption frequency,
$\nu_{a} \approx 2 \times 10^{11}$~Hz. For $\nu_a < \nu_m$, as is
the case here, the self absorption coefficient scales as
$\alpha_{\nu}\propto\nu^{-5/3}$ \citep[see][equation (6.50)]{RL79},
thus $\tau_{\nu}/\tau_{\nu_{m}} =
\alpha_{\nu}/\alpha_{\nu_{m}}=\left (\nu/\nu_{m}\right )^{-5/3}$.
Here, $\tau_{\nu}\propto\alpha_{\nu}$ is the optical depth,
$\tau_{\nu_{m}}\equiv\tau_{\nu}(\nu=\nu_{m})=\alpha_{\nu_{m}}R^{\prime}$,
and $R^{\prime}=R/\Gamma$ is the co-moving size of the synchrotron
emitting region. In calculating $\alpha_{\nu_m}$ we use equation
(6.53) in \citet{RL79}, which assumes power law distribution of
electrons above $\gamma_{e}$. We use power law index $p=2$, although
the result is found not to be sensitive to neither the power law
distribution assumption or the power law index; a similar result is
obtained if the electrons assume a thermal distribution.  Since, by
definition, $\tau_{\nu_{a}}\equiv\tau_{\nu}(\nu=\nu_{a})=1$, the
synchrotron self absorption frequency $\nu_{a}$ is
\beq
\frac{\nu_{a}^{\rm ob}}{\nu_{m}^{ob}}=2.46\times 10^{-6}
N_{e}^{3/5}R^{-6/5}B^{-3/5}\gamma_{e}^{-3} \simeq {1 \over 4}~.
\label{absorption-peak-frequency2}
\eeq
Here, the observed value of the self absorption frequency
$\nu_{a}^{\rm ob} \approx 2\times 10^{11}$~Hz is used.

As a fourth condition we use the assumption of ballistic expansion of
the source, similar to \citet{Zauderer+11}. We consider constant bulk
expansion velocity of the source $\beta =\left (1-\Gamma^{-2}\right
)^{1/2}$ and head-on emission. At redshift $z=0.354$, an observed time
$\Delta t^{\rm ob}=5$~days corresponds to source frame time $\Delta
t^{\rm ob}/(1+z)=3.7$~days.  Due to relativistic time compression, the
actual time that the source would have expanded is $[\Delta
t^{\rm ob}/(1+z)]/(1-\beta)$. Thus, the emission radius is related to the
observed time by
\begin{eqnarray}
R=\frac{\Delta t^{\rm ob}}{(1+z)}\cdot\frac{\beta c}{(1-\beta)}
\label{ballistic-expansion}
\end{eqnarray}

\begin{figure}[ht]
\centerline{\includegraphics[width=12cm]{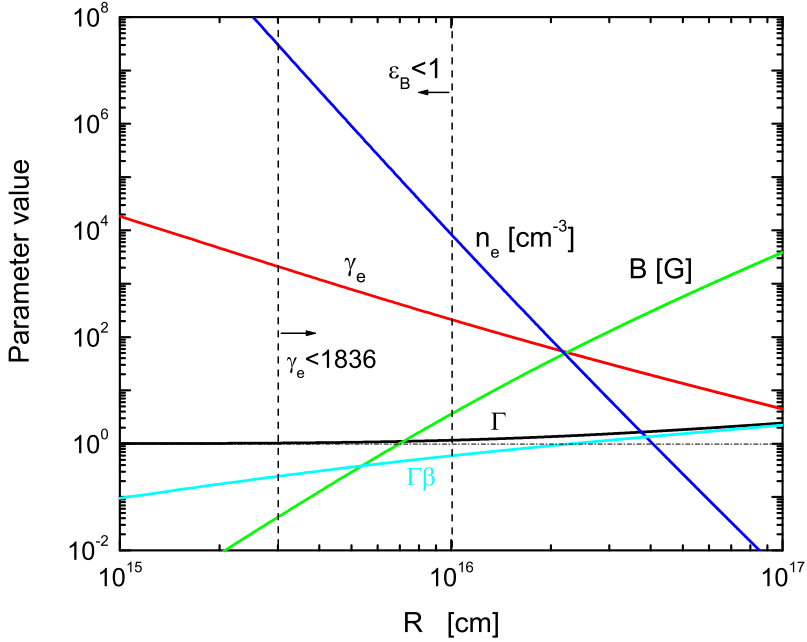}}
\caption{Dependence of the free model parameters: the bulk Lorentz
  factor, $\Gamma$ (black), the bulk momentum, $\Gamma\beta$ (cyan),
  the magnetic field, $B$ (green), the typical Lorentz factor of
  electrons, $\gamma_{e}$ (red) and the number density of electrons
  $n_{e}$ (in the observer's frame) (blue) on the emission radius of the radio photons, $R$,
  under the assumption that the radio photons originating from
  synchrotron emission.
} \nonumber
\label{fig:1}
\end{figure}

The four constraints derived in equations (\ref{totalflux}) --
(\ref{ballistic-expansion}) are insufficient to fully determine the
values of the five free model parameters. We therefore choose the
source size $R$ at $\Delta t^{\rm ob} = 5$~days as a free variable,
and determine the values of the other four free parameters. The
results of our calculation are shown in Figure \ref{fig:1}. We
present the bulk Lorentz factor $\Gamma$ (black), bulk momentum
$\Gamma\beta$ (cyan), magnetic field $B$ (green), characteristic
electron Lorentz factor $\gamma_{e}$ (red) and the number density of
the radiating particles, in the observer's frame, $n_{e}$
(blue).\footnote{The number density is calculated using $N_e =
  (4\pi/3) n_e R^3$. Note that for $\Gamma \gtrsim 1$, the number
  density in the comoving frame is similar to the number density in
  the observer's frame.}

In order to constrain the allowed parameter space region, we use two
additional assumptions: (1) In Fermi-type acceleration, the typical
Lorentz factor of the energetic electrons $\gamma_{e} \leq
m_{p}/m_{e}=1836$ in the rest frame of the plasma. This
  assumption follows an equipartition assumption between the energy
  given to the accelerated electrons and protons; (2) As the
emission radius is large, the magnetic field must be produced at the
shock front. Thus, the ratio of magnetic energy density,
$B^{2}/8\pi$ to the photon energy density, $L/(4\pi
R^{2}\Gamma^{2}c)$ (often denoted by $\epsilon_{B}$), is smaller
than unity. These constraints which are commonly used in modeling,
e.g., emission from
  GRBs, are shown by the dashed lines in Figure \ref{fig:1}.

After adding these two constraints, we conclude that the radio
emission zone at early times fulfills the following conditions: {\it
  (i)} The emission radius is in the range $3.0\times 10^{15}~{\rm
  cm}\lesssim R \lesssim 1.0\times 10^{16}$~cm.  {\it (ii)} The
outflow is trans-relativistic, with $1.04\lesssim \Gamma\lesssim
1.2$, and $0.24\lesssim\beta \lesssim0.55$. {\it (iii)} The magnetic
field is poorly constrained by the data, and could range from
$5.0\times 10^{-2}$~G to $7.0$~G. {\it (iv)} The electrons are hot,
with minimum random Lorentz factor exceeding $\gamma_{e} \gtrsim
150$. {\it (v)} The emitting region is dense: $3.0\times 10^{3}~{\rm
cm}^{-3}< n_{e}< 3.0\times 10^{7}~{\rm cm}^{-3}$. This result likely
excludes shocked external ISM material as the source of the radio
emission, as the typical densities of the ISM, $\approx 1-10~{\rm
cm}^{-3}$ \citep{Baganoff+03}, even after compressed by
mild-relativistic shock waves -  are much lower than these values.
{\it (vi)} The total number of radiating particles is at the range
$1.0\times 10^{52}< N_{e}<3.4\times 10^{54}$, with likely value of
$\sim 10^{53}$.  These results are consistent with the results
derived by \citet{Zauderer+11}.


\subsection{Interpretation of early X-ray emission}
\label{sec:X-rays}

Our underlying assumption is that the origin of the X-ray emission at
early times is mainly due to inverse Compton (IC) scattering of the
synchrotron radio photons by relativistic electrons. As we show here,
these electrons must be located in a different region than the
radio-emitting electrons. Following the treatment by \citet{PL12}, two
constraints can be put by the data: (I) the ratio of the IC and
synchrotron peak frequencies is given by
\begin{eqnarray}
\frac{\nu_{\rm peak, IC}^{\rm ob}}{\nu_{\rm peak, syn}^{\rm ob}}=\frac{4}{3}\left
(\gamma_{\rm e[IC]}\Gamma\right )^{2} \simeq {2 \times 10^{18} \over 8
  \times 10^{11}},
\label{nuICpeak}
\end{eqnarray}
and, (II) the ratio of IC to synchrotron peak fluxes is
\begin{eqnarray}
 \frac{F_{\nu,\rm IC}}{F_{\nu,\rm syn}} = { (\nu F_{\nu, peak, IC} /
   \nu_{\rm peak, IC}) \over (\nu F_{\nu, peak, syn} / \nu_{\rm peak,
     syn})} = n_{e [IC]}r\sigma_{\rm T} \simeq 4 \times 10^{-3}.
\label{luminosityICpeak}
\end{eqnarray}
Here, $\gamma_{e[IC]}$ and $n_{e[IC]}$ are the typical Lorentz
factor and number density of electrons (in the observer's frame)
that emit the IC photons, and $r$ and $\Gamma$ are the typical size
and the bulk Lorentz factor of the IC emission region. In estimating
the ratio of the peak frequencies and fluxes, we used $\nu_{\rm
peak,
  IC}^{\rm ob} \simeq 2\times 10^{18}$~Hz, $\nu_{\rm peak, syn}^{\rm
  ob} \simeq 8 \times 10^{11}$~Hz, $\nu F_{\nu, peak, IC} \approx 3
\times 10^{47}~{\rm erg~s^{-1}}$, and $\nu F_{\nu, peak, syn}
\approx 3 \times 10^{43}~{\rm erg~s^{-1}}$. Similar to the analysis
of
  the radio data, spherical explosion was assumed here as well.

The rapid variability observed in X-rays on a time scale $\delta
t^{\rm ob} \sim 100$~s constraints the size of the emitting region, $r$.
While the time during which this variability is observed does not
correspond directly to five~days, we use it here as an order of
magnitude estimate. This variability implies a relation between the
emission radius of the X-rays and the bulk Lorentz factor,
\begin{eqnarray}
r=\left (\frac{\delta t^{ob}}{1+z}\right )\frac{\beta c}{1-\beta}\approx
2\Gamma^{2}c\frac{\delta t^{ob}}{1+z}
\label{eq:Xray-radius}
\end{eqnarray}

\begin{figure}[ht]
\centerline{\includegraphics[width=12cm]{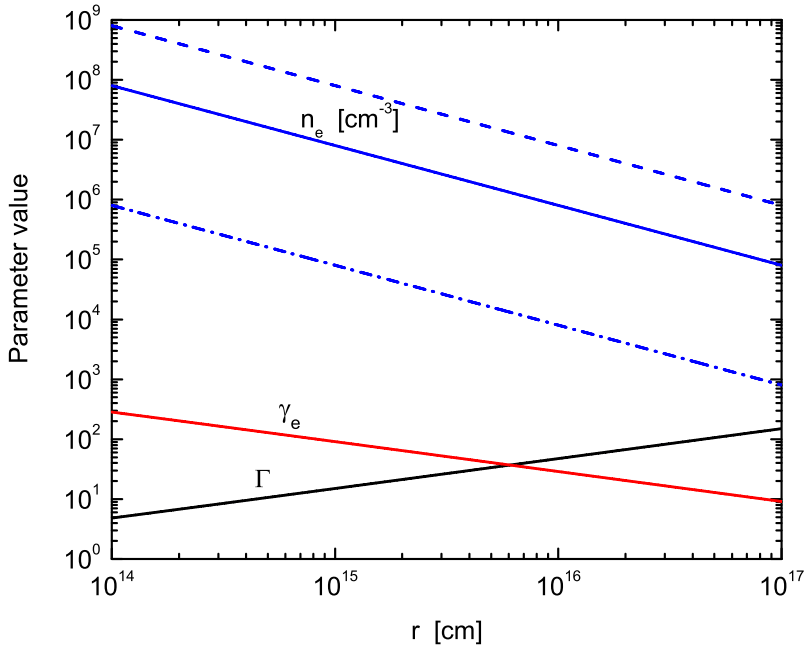}}
\caption{Dependence of the free model parameters of the X-ray emitting
  region: the bulk Lorentz factor, $\Gamma$ (black) the typical
  Lorentz factor of electrons, $\gamma_{e}$ (red) and the number
  density of electrons in the observer's frame, $n_{e}$
  (blue) on the radius of source, $r$. The solid, the dashed
    and the dash-dotted lines represent the averaged X-ray luminosity
    $3 \times 10^{47}~{\rm erg~s^{-1}}$, the maximum luminosity $\sim
    3 \times 10^{48} \, {\rm erg \, s^{-1}}$ and the minimum
    luminosity $\sim 3\times 10^{45} \, {\rm erg \, s^{-1}}$,
    respectively. We assume that IC scattering of the radio photons
  are the main source of X-rays.  }
 \label{fig:2}
\end{figure}

Equations (\ref{nuICpeak}), (\ref{luminosityICpeak}) and
(\ref{eq:Xray-radius}) exhibit three restrictive conditions provided
by the observed data. As there are four free model parameters,
$\Gamma$, $\gamma_{e[IC]}$, $n_{e[IC]}$ (or $N_{e}$) and $r$, a full
solution cannot be obtained. However, we can apply a similar method
to the one used in \S\ref{sec:radio1}, namely take $r$ as a free
parameter and obtain the values of the other three unknowns. The
results of this analysis are presented in Figure \ref{fig:2}. In
this figure, we present the values of $\Gamma$ (black),
$\gamma_{e[IC]}$ (red) and $n_{e[IC]}$ (blue) as a function of $r$,
where $\delta t^{ob} = 100$~s is considered.

The X-ray luminosity varies in the range $\sim 3\times 10^{45}
  \, {\rm erg \, s^{-1}}$ --- $\sim 3 \times 10^{48} \, {\rm erg \,
    s^{-1}}$. This uncertainty leads to an uncertainty in the number
  density (or the total number of) electrons. However, the derived
  values of the bulk Lorentz factor and the typical Lorentz factor of
  electrons (Equations \ref{nuICpeak} and \ref{eq:Xray-radius}) are
  not affected by this uncertainty. In the results presented in Figure
  \ref{fig:2}, we thus present three values of the number density (as
  a function of $r$), obtained by taking different observed fluxes of
  the X-rays: The solid blue line corresponds to the average X-ray
  flux, $3 \times 10^{47}~{\rm erg~s^{-1}}$, while the dash and
  dash-dotted lines correspond to the maximum and minimum observed
  X-ray flux, respectively.

From the results of Figure \ref{fig:2} one can put several
constraints on the X-ray emission zone: First, there is no strong
constraint on the size of the emitting region, $r$. Any value in the
range $10^{15}~{\rm cm}< r <10^{16}~{\rm cm}$, which is compatible
with the size of the emission zone of the synchrotron photons, is
acceptable. The main constraint originates from equation
(\ref{eq:Xray-radius}), as large radius implies large bulk Lorentz
factor, since $\Gamma \simeq r^{1/2} / (2.1 \times 10^6)$. Thus,
value of $r \approx 10^{16}$~cm imply $\Gamma \approx 50$, much
larger than the value obtained for the radio emission zone. Even
value of $r = 10^{15}$~cm implies $\Gamma \approx 15$, inconsistent
with the finding for the radio emitting zone.  Thus, while the size
of the radio and X-ray emitting regions can be comparable, the X-ray
emitting region must propagate at much larger Lorentz factor than
the radio emitting region. Second, the IC emitting electrons are not
as hot as the radio emitting electrons. For $r \gtrsim 10^{15}$~cm,
$\Gamma \gtrsim 15$, which, using equation (\ref{nuICpeak}) imply
$\gamma_{e[IC]}\lesssim 100$. Third, the number density of the
radiating electrons, $10^4 {\rm \, cm^{-3}} \lesssim n \lesssim 10^8
{\rm \,
  cm^{-3}}$ is comparable to the number density of the radio emitting
particles, and is thus likely too high to be explained by compression
of the external material. A similar conclusion holds for the total
number of the radiating particles.

Thus, we conclude that while the two emission regions can have a
comparable size, the X-ray emission zone must have a much larger
Lorentz factor than the radio emission zone; moreover, the electrons
in the X-ray emitting region must be colder than the electrons that
emit at radio frequencies. Alternatively, the bulk motion may be
similar, but only if the X-ray emitting region is at much smaller
distance, $10^{12.5}$~cm. These results are summarized in Table
\ref{tab01}. The numbers for the X-ray emission region in Table 1
are obtained by taking the average X-ray luminosity, $L_{x}=3\times
10^{47} \, {\rm erg \, s^{-1}}$.

\begin{table}[htbp]
\caption{Summary of key properties of both radio and X-ray emission
zones of Swift J164449.3+573451}
\begin{tabular}
{cccccc} \hline\hline \raisebox{-1.50ex}[0cm][0cm]{}&
 ~~$R\,{\rm [cm]}$~~  & ~~$\Gamma$~~ & ~~$\gamma_{e}$~~ & ~~$n_{e} \,{\rm [cm^{-3}]}$~~ & ~~$N_{e}$~~\\
\hline \raisebox{-0.50ex}[0cm][0cm] {Radio emission} {}& & & & {} &
\\\raisebox{-0.50ex}[0cm][0cm] {zone}
& $3 \times 10^{15}-10^{16}$ & $\lesssim 1.2$ & $150-2000$ & $10^{3.5}- 10^{7.5}$ & $10^{52}-10^{54.5}$\\
\hline \raisebox{-0.50ex}[0cm][0cm] {X-ray emission} {}& & & & {} &
\\\raisebox{-0.50ex}[0cm][0cm] {zone}
& $10^{15}-10^{16}$ & $15 - 50$ & $30 - 100$ & $10^{5}-10^{8}$ & $10^{52}-10^{54.5}$\\
\raisebox{-0.50ex}[0cm][0cm] {(alternative)}
& $10^{12.5}$ & $\lesssim 1.2$ & $\sim 1000$ & $10^{9.5}$ & $10^{47.5}$\\
\hline
\label{tab01}
\end{tabular}
~~~~{\small Notes. --- In the table, the model parameters $R$:
source radius, $\Gamma$: bulk Lorentz factor, $\gamma_{e}$: typical
Lorentz factor of relativistic electrons, $n_{e}$: number density of
emitting electrons, and $N_{e}$: total number of electrons in the
observer's frame.}
\end{table}

Similarly, although order-of-magnitude variations in brightness are
seen in the X-rays, the detailed radio light curve does not reveal the
coincident variations that would be expected for SSC
\citep{Zauderer+11, Levan+11}. We therefore conclude that the X-ray
emission must originate from a region separated than the radio
emission region.

\begin{figure}[ht]
\centerline{\includegraphics[width=12cm]{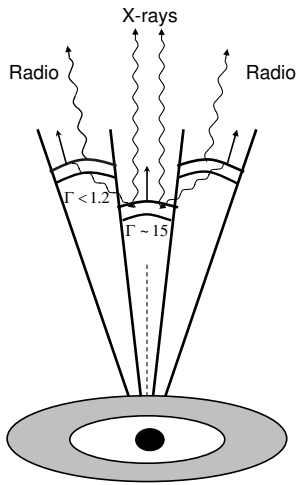}}
\caption{ Sketch of geometrical configuration and emission regions for
  TDE Swift 1644+57. The inner jet has a Lorentz factor $\Gamma
  \approx 15$, and is responsible for the early times X-ray
  emission. The outer jet has Lorentz factor $\lesssim 1.2$, and is
  the source of the early times radio emission. As the inner jet
  propagates through the ISM it collects material and
  cools. Synchrotron emission from the collected material is the
  source of the re-brightening seen in radio frequencies after
  $\approx 30$~days.
}
\label{fig:3}
\end{figure}

While observations made in the first few days cannot discriminate
between the two alternatives for the X-ray emitting region, we show in
\S\ref{sec:late_times} below that the re-brightening observed at
radio flux after tens of days can be naturally explained as resulting
from material collected by the X-ray emitting plasma, provided that it
travels at $\Gamma \approx 10 - 20$. Thus, the first model, in which
the X-ray emission region is separated from the radio emission region
{\it in velocity space} is preferred. This leads us to a proposed {\it
  two component jet}, in which the fast, X-ray emitting material, is
surrounded by a slower, radio emitting material. A cartoon
demonstrating this model is shown in Figure \ref{fig:3}.

\subsection{Constraints on GeV emission}
\label{sec:GeV}

In the framework of the two-component jet model proposed here, a
strong $\sim$~GeV flux is produced. This results from inverse Compton
scattering of the X-ray photons produced in the inner jet by energetic
electrons in the same region.  Using the outflow parameters derived
above (see Table \ref{tab01}), we find that the X-ray photons (of
characteristic observed energy $ \sim 10$~keV) will be upscattered to
observed energies $\sim\gamma_{e}^{2} h\nu_{x}\sim 100$ MeV.
Accordingly, the luminosity of this sub-GeV emission is expected to be
$L_{\rm GeV}/L_{x}= Y = n_{e}\sigma_{T} r \gamma_{e}^{2} \gtrsim 10$,
where $Y$ is Compton parameter. Using the observed X-ray luminosity
$L_{x}\sim 3\times 10^{47}~{\rm erg/s}$, one can naively expect
$L_{\rm GeV} \gtrsim 10^{48}~{\rm erg/s}$, which exceeds the limits
set by Fermi \citep{Burrows+11} by about two orders of magnitude.

This GeV emission, however, is strongly suppressed by annihilating
with the X-ray photons, and cannot, therefore, be detected. Photons at
$\sim 1$~GeV (observed energy) will annihilate with photons at
observed energies in the X-ray band, of $\gtrsim 50$~keV (assuming
bulk Lorentz factor $ \Gamma \sim 15$). The (comoving) number density
of these X-ray photons is $n_x' \approx L_x / (4 \pi r^2 \Gamma^2 c
\langle \epsilon' \rangle) \approx 7 \times 10^{11}~{\rm\,cm^{-3}}$,
where we have assumed $r \simeq 10^{15}$~cm, $\Gamma \approx 15$ and
$\langle \epsilon' \rangle \approx 3$~keV is the comoving energy of
the X-ray photons.  The optical depth for pair production is thus
$\tau_{\gamma\gamma} \approx (r/\Gamma) n_x' {\sigma}_{\gamma\gamma}=
5$, where the cross section for $\gamma-\gamma$ annihilation is
${\sigma}_{\gamma\gamma}\sim 10^{-25}~{\rm cm}^{2}$.  This attenuation
can explain the lack of detection of GeV photons by Fermi
\citep{Burrows+11}.

We further point that $\approx 100$~MeV photons annihilate
  with photons at $\sim 500$~KeV, whose flux may be lower than that of
  50~keV photons. However, the Fermi upper limits are obtained on the
  integrated flux between 100~MeV and 10~GeV; Moreover, as the XRT
  bandwidth is limited to below 150~keV, no observational constraints
  exist on the number density of photons, hence on the optical depth
  for annihilation at these energies.


\section{Late time radio evolution}
\label{sec:late_times}

The two component jet model presented above has a distinct
prediction. As the plasma expands into the interstellar material
(ISM), it collects material, slows and cools. This situation is
similar to the afterglow phase in GRBs \citep[e.g.,][and references
  therein]{ZM09, GP12, vEM12, vE13}. Thus, using similar assumptions,
one can predict the late time synchrotron emission from the
decelerating plasma. One notable difference from GRB afterglow,
though, is that the X-ray emitting plasma is mildly relativistic,
while the radio emitting plasma is trans-relativistic. Thus, when
calculating the dynamics, one cannot rely on the ultra-relativistic
scheme \citep{BM76}, but has to consider the transition to the
Newtonian regime.

\subsection{Dynamics and radiation from an expanding jet}
\label{sec:dynamics}

The dynamics of plasma expanding through the ISM is well studied in
the literature \citep{BM76, KP97, CD99, Piran99, Huang+99, vPKW00,
  Peer12, Nava+12}.  Here we briefly review the basic theory, which we
then utilize in calculating the expected late time radio emission from
TDE Swift 1644+57.

We assume that by the relevant times (tens of days), the reverse shock
had crossed the plasma, and thus only the forward shock exists. The
evolution of the bulk Lorentz factor of the expanding plasma is given
by \citep{Peer12}
\beq
\frac{{\rm d}\Gamma}{{\rm d}m}=-\frac{\hat{\gamma}\left
  (\Gamma^{2}-1\right )-\left (\hat{\gamma}-1\right )\Gamma\beta^{2}}
     {M+\epsilon m+\left (1-\epsilon\right )m\left
       [2\hat{\gamma}\Gamma -\left (\hat{\gamma}-1\right )\left
         (1+\Gamma^{-2}\right )\right ]}.
\label{dynamicalevolution} \eeq Here, $M$ is the mass of the ejected
matter, $m$ is the mass of the collected ISM, $\Gamma$ is the bulk
Lorentz factor of the flow, $\hat{\gamma}$ is the adiabatic index
and $\epsilon$ is the fraction of the shock-generated thermal energy
that is radiated ($\epsilon =0$ in the adiabatic case and $\epsilon
=1$ in the radiative case).  Note that equation
(\ref{dynamicalevolution}) holds for any value of $\Gamma$, both in
the ultra-relativistic ($\Gamma\gg 1$) and the sub-relativistic
($\beta\ll 1$) limits.

Under the assumption of constant ISM density, the collected ISM mass is
related to the distance ${\rm d}R$ and observed time ${\rm d}t$ via
\beq
\ba{lcl}
{\rm d}m & = & 4\pi R^{2} n_{\rm ISM}m_{\rm p}{\rm d}R ; \\
{\rm d}R & = & \Gamma\beta c\left (\Gamma +\Gamma\beta\right ){\rm
  d} t=\frac{\beta}{1-\beta}c{\rm d}t,
\ea \label{eq:cm} \eeq where $m_{p}$ is the proton's rest mass and
$n_{\rm ISM}$ is the ISM density. In deriving the second line in
equation (\ref{eq:cm}), we have explicitly assumed that the observed
photons are emitted from a plasma that propagates towards the
observer. A more comprehensive calculation which considers the
integrated emission from different angles to the line of sight
results in a similar solution, up to a numerical factor of a few
\citep{Waxman97, PW06}.

In order to predict the synchrotron emission, one needs to calculate
the magnetic field, $B$ and characteristic electron's Lorentz factor,
$\gamma_{\rm el}$. This calculation is done as follows.  By solving
the shock jump conditions, one gets the energy density behind the
shock \citep{BM76},
\beq
u_{2}=\left (\Gamma -1\right )\frac{\hat{\gamma}\Gamma
  +1}{\hat{\gamma}-1}n_{\rm ISM}m_{\rm p}c^{2}~.
\label{energy-density}
\eeq
Equation (\ref{energy-density}) is exact
for any velocity, including both the ultra-relativistic and the
Newtonian limits.  A useful approximation for the adiabatic index is
$\hat{\gamma}=( 4 \Gamma + 1)/(3 \Gamma)$
\citep[e.g.,][]{Dai+99}\footnote{A more accurate
  formula appears in \citet{Peer12}.}.  In the ultra-relativistic
limit, $\Gamma \gg 1$, $\hat{\gamma}=4/3$ and equation
(\ref{energy-density}) takes the form $u_{2}\approx
4\Gamma^{2}n_{\rm
  ISM}m_{\rm p}c^{2}$; While in the Newtonian limit, $\beta\ll 1$ and
$\hat{\gamma}=5/3$, equation (\ref{energy-density}) becomes
$u_{2}\approx 2\beta^{2}n_{\rm ISM}m_{\rm p}c^{2}$.

The shock-generated magnetic field assumes to carry a fraction
$\epsilon_{\rm B}$ of the post-shock thermal energy,
$B^{2}/8\pi=\epsilon_{\rm B} u_2$,
\begin{eqnarray}
B = \left \{ \begin{array}{ll} (32\pi\epsilon_{\rm B}n_{\rm ISM}m_{\rm
    p}c^{2})^{1/2}\Gamma & ({\rm relativistic~limit}), \\
\left (16\pi\epsilon_{\rm B}n_{\rm ISM}m_{\rm p}c^{2}\right
)^{1/2}\beta & ({\rm Newtonian~limit}).
\end{array}
\right .
\label{magneticfield}
\end{eqnarray}
Similarly, a constant fraction $\epsilon_{\rm e}$ of the post-shock
thermal energy is assumed to be carried by energetic electrons,
resulting in $\gamma_{\rm el}m_{\rm e}c^{2} = \epsilon_{\rm
e}(u_{2}/n_2)$, where the number density in the shocked region is
$n_{2}=n_{\rm ISM}(\hat{\gamma}\Gamma+1)/(\hat{\gamma}-1)$.
This leads to
\beq
\gamma_{\rm el}=\left \{ \begin{array}{ll} \epsilon_{\rm e}\Gamma\left
  (\frac{m_{\rm p}}{m_{\rm e}}\right ) & ({\rm
    relativistic~limit}), \\ \epsilon_{\rm e}\frac{\beta^{2}}{2}\left
  (\frac{m_{\rm p}}{m_{\rm e}}\right ) & ({\rm
    Newtonian~limit}).
\end{array}
\right .
\label{electronenergy}
\eeq
In order to calculate the observed peak frequency and flux, we
discriminate between two cases:

(i) In the {\it optically thin} emission, $\nu_{a}< \nu_{m}$, i.e.,
$\tau_{\nu_{m}}<1$, the peak synchrotron frequency is $\nu_p^{\rm
ob} = \nu_m^{\rm ob} = (3/4\pi) (q_{e} B /m_e c) \gamma_{\rm el}^2
\Gamma$ and the peak flux $F_{\nu_{p}} = F_{\nu_{m}}\approx (N_e / 4
\pi d_L^2) (2 \sigma_T m_e c^2 B \Gamma / 9 q_{e})$
\citep[e.g.,][]{SPN98}. Here, $d_L$ is the luminosity distance and
$N_e \propto R^3$ is the total number of radiating electrons
(originating both at the explosion, as well as collected ISM). In
the limit where the total number of swept-up ISM material is much
larger than the original ejected material, analytic scaling laws can
be obtained in the ultra-relativistic and Newtonian limits. In the
relativistic regime, $\Gamma \gg 1$, $\Gamma \propto t^{-3/8}$ and
$R \propto t^{1/4}$, and thus $\nu_p^{\rm ob} \propto t^{-3/2}$ and
$F_{\nu_p} \propto t^0$. On the other hand, in the Newtonian limit,
$\beta \propto t^{-3/5}$ and $R \propto t^{2/5}$, one finds
$\nu_p^{\rm ob} \propto t^{-3}$ and $F_{\nu_p} \propto t^{3/5}$.

The temporal evolution of the self absorption frequency is
calculated using the relation between $\nu_a$ and $\nu_m$ (Equation
\ref{absorption-peak-frequency2}), the scaling laws of $B$ and
$\gamma_{el}$ derived in Equations \ref{magneticfield} and
\ref{electronenergy}, the relation $N_e \propto R^3$ and the scaling
laws of $R$ and $\Gamma$.  Using the results of Equation
\ref{absorption-peak-frequency2}, one can write $\nu_a \propto
\nu_p^{\rm ob} N_e^{3/5} R^{-6/5} B^{-3/5} \gamma_{el}^{-3}$. In the
ultra-relativistic case, $B \propto \Gamma$, $\gamma_{el} \propto
\Gamma$, $\Gamma \propto t^{-3/8}$ and $R \propto t^{1/4}$, and
therefore $\nu_a \propto t^0$, namely time-independent.  In the
other extreme of Newtonian motion, $B \propto \beta$, $\gamma_{el}
\propto \beta^2$, $\beta \propto t^{-3/5}$ and $R \propto t^{2/5}$,
leading to an {\it increase} of the self absorption frequency with
time, $\nu_a \propto t^{6/5}$.

(ii) In the {\it optically thick} regime, $\nu_{a}> \nu_{m}$, i.e.,
$\tau_{\nu_{m}}>1$, the observed peak frequency $\nu_{p}$ and the peak
flux $F_{\nu_{p}}$ are at the self absorption frequency, $\nu_a$. For
a power law distribution of electrons above $\gamma_{\rm el}$ with
power law index $p$, these are given by
\begin{equation}
\ba{rcl}
\nu_{p}^{\rm ob} =\nu_a^{\rm ob} = \nu_{m}^{\rm ob} \tau_{\nu_{m}}^{2/(p+4)} & = &
\frac{3}{4\pi}\frac{q_{e}B}{m_{\rm
    e}c}\gamma_{\rm el}^{2}\Gamma\tau_{\nu_{m}}^{2/(p+4)}~, \\
F_{\nu_{p}}=F_{\nu_{m}}\left (\frac{\nu_{a}}{\nu_{m}}\right
)^{-\frac{p-1}{2}}& =& F_{\nu_{m}}\tau_{\nu_{m}}^{-\frac{p-1}{p+4}}
\approx \frac{N_{\rm e}}{\rm 4\pi
  d_{L}^{2}}\frac{2}{9}\frac{\sigma_{\rm T}m_{\rm
    e}c^{2}}{q_{e}}B\Gamma\tau_{\nu_{m}}^{-\frac{p-1}{p+4}}~.
\ea
\label{peakfrequency2}
\end{equation}

The optical depth at $\nu_m$, $\tau_{\nu_m}$ is calculated as
follows. We first point out that the comoving number density of the
electrons in the shocked plasma frame is $n_{2} \simeq 4\Gamma n_{\rm
  ISM}$.\footnote{The equation relating the downstream ($n_2)$ and
  upstream ($n_1$) number densities, $n_2 = (\hat \gamma \Gamma + 1)
  n_1/(\hat \gamma -1)$ holds for any strong shock, at any velocity,
  both in the relativistic and Newtonian regimes.} Using equation
  (6.52) in \citet{RL79}, we find that $\tau_{\nu_{m}}=\alpha_{\nu_{m}}
(R/\Gamma) = f(p) n_{\rm ISM} B^{-1} \gamma_{\rm el}^{-5} R$ \citep[see
  also][]{PW04}. Here, $f(p) = (p-1) (8 \pi \sqrt{3} q_e /9) 2^{p/2}
\Gamma((2p+2)/12) \Gamma((3p+22)/12)$ is a function of the power law
index, $p$ of the accelerated electrons above $\gamma_{\rm el}$, and
$\Gamma(x)$ is $\Gamma$ function of argument $x$.  Using this result
in equation (\ref{peakfrequency2}), one can obtain analytic scaling
laws for the temporal evolution of $\nu_p^{\rm ob}$ and $F_{\nu_p}$
in both the ultra-relativistic and Newtonian limits, in a similar
way to the derived expressions in the optically thin limit above. In
the relativistic regime, $\nu_p^{\rm ob} \propto t^{-(3p+2)/2(p+4)}
\propto t^{-2/3} \, (p=2)$ and $F_{\nu_p} \propto t^{5(1-p)/2(p+4)}
\propto t^{-5/12} \, (p=2)$. In the Newtonian limit, one finds
$\nu_p^{\rm ob} \propto t^{(2-3p)/(p+4)} \propto t^{-2/3} \, (p=2)$
and $F_{\nu_p} \propto t^{(47-32p)/5(p+4)} \propto t^{-17/30} \,
(p=2)$.

The scaling laws presented above are derived under the
  assumption that the origin of the magnetic field and the source of
  energy of energetic particles originates from randomization of the
  kinetic energy at the shock front. As such, these scaling laws are
  valid for the collected ISM plasma. However, we argue that similar
  scaling laws hold for the plasma originally ejected in the jet, at
  least at late times (after a few days) when the ejecta radii is much
  larger than its original radii, and the motion becomes self similar.

The ejected material and the collected ISM are separated by a
  contact discontinuity. By definition, the energy densities at both
  sides of the discontinuity are the same. Thus, using the common
  assumption that constant fractions of the energy density are
  used in generating magnetic fields and accelerating particles, one
  must conclude similar values at both sides of the contact
  discontinuity. Moreover, the contact discontinuity is unstable,
  implying a mixture of materials at both its sides
  \citep{DM14}. Thus, even if the values of the physical parameters
  were initially different at both sides of the discontinuity, we
  expect them to average out at late times.

\subsection{Radiative cooling of the electrons}
\label{sec:cooling}

In the calculation presented in \S\ref{sec:dynamics} (in particular,
equation (\ref{electronenergy}) above), we considered heating of the
ISM plasma as it crosses the shock front. Once the ISM particles
cross the shock front, they lose their energy by radiative cooling.
Hence, their typical Lorentz factor $\gamma_{\rm el}$ decreases with
time. Similarly, the original ejected material from the tidally
disrupted star cools with time.

Within the context of our double jet model, this cooling has a more
pronounced effect on the lightcurve originating from the outer jet,
that is responsible for the early time radio emission. This is because
this region is composed of a dense plasma, that propagates at
trans-relativistic velocities (see Table \ref{tab01}). The
contribution of the swept-up ISM material to the emission during the
first $\sim$~month from this region is therefore minor, as opposed to the
contribution of the collected ISM to the emission from the inner
jet. This can be seen by noting that during Newtonian expansion, $R
\propto t^{2/5}$. Thus, between five and thirty days, the radius of
the radio emitting region is increased by a factor $\approx 2$, and
therefore cannot exceed $\approx 2 \times 10^{16}$~cm. As a result, unless
the ISM density is much larger than $\sim 10^3 \, \rm{cm^{-3}}$,
contribution from the collected ISM to the emission from the outer jet
is subdominant.

Although the total number of collected ISM particle within the first
month is considerably smaller than the number of particle that
existed initially in the outer jet, the collection of the ISM
material caused the jet to slow. \textbf{Our underlying assumption
is that by 5 days, the outer jet have already reached its
self-similar phase, whose scaling laws we derived above. The slowing
down and expansion of the jet} leads to a decrease in the energy
density, hence of both the photon and magnetic field with time.
This, in turn, modifies the electrons cooling rate. The cooling of
the electrons, in turn, leads to a decay of the radio emission (from
the electrons that initially exist in the outer jet). Here we
calculate the temporal evolution of the electron's Lorentz factor,
resulting from this radiative decay. \textbf{We stress that in the
framework of our model, both the initially ejected outer jet
material and collected ISM material radiate, under similar
conditions (same magnetic field, etc.). While this may only be a
crude approximation \citep[e.g.,][]{DM14}, the results obtained are
consistent with the data, implying that this assumption may be
valid.}


The radio emitting electrons cool due to synchrotron and inverse
Compton scattering. The radiated power from a relativistic electron
with Lorentz factor $\gamma_{\rm el}$ is $P = P_{\rm syn}+P_{\rm
  IC}=(4/3)c \sigma_{T}\left (U_{B}+U_{\rm ph}\right )\gamma_{\rm
  el}^{2}$, where $U_{B}$ and $U_{\rm ph}$ are the magnetic and
radiative field energy densities, respectively.

Both $U_{B}$ and $U_{\rm ph}$ decrease with radius (and time). As
discussed in \S\ref{sec:dynamics} above, during the Newtonian
expansion phase $B \propto \beta \propto t^{-3/5}$ (see equation
(\ref{magneticfield})), and thus $U_B \propto t^{-6/5}$. Similarly,
$U_{\rm ph}=L/(4\pi R^{2}\Gamma^{2}c)$, where the luminosity is $L
\propto N_{e}\gamma_{\rm el}^{2}B^{2}\Gamma^{2} \propto \beta^6
\propto t^{-18/5} $ (assuming that most radiative particles are from
the original ejecta). This leads to $U_{\rm ph} \propto t^{-22/5}$,
namely the IC component decreases much faster than the synchrotron
component.

We can therefore write $P \approx (4/3) c \sigma_T U_{B,0}
(t/t_0)^{-6/5} \gamma_{\rm el}^2$, where $U_{B,0}$ is the magnetic
field energy density at fiducial time, $t_0$ which is taken in the
calculations below to be five days (observed time). The characteristic
particles Lorentz factor at any given time $t>t_0$ is given by
\begin{eqnarray}
\gamma_{\rm el}(t) =\frac{\gamma_{\rm el,0}}{1+{20 \over 3}{\sigma_T \over m_e
    c}U_{B,{0}}\gamma_{\rm el,0}t_{0}\left [1-\left (t/t_{0}\right
    )^{-1/5}\right ]}~,
\label{gammatrappingtime}
\end{eqnarray}
where $\gamma_{\rm el,0}$ is the electron's Lorentz factor at $t_0$. The
temporal evolution of the electron's Lorentz factor and the decay
law of the magnetic field allow calculation of the late time
evolution of both the peak radio frequency and peak radio flux from
the outer jet region.

\subsection{Lateral expansion}
\label{sec:lateral}

The temporal evolution of the observed peak frequency and peak flux
were derived above under the assumption of spherical expansion. In a
jetted outflow, initially the sideways expansion can be neglected;
however, once the outflow decelerates to $\Gamma \approx
\theta_j^{-1}$ (where $\theta_j$ is the initial jet opening angle),
lateral expansion becomes significant. The flow expands sideways
during its trans-relativistic phase, which lasts a typical observed
time of few - few hundred days before asymptoting to a spherical,
Newtonian expansion described by the Sedov-Taylor solutions
discussed above \citep{Livio+00, vEMWK11, WWF11}.

In a jetted outflow with $\theta_j \ll 1$, the true jet energy is $E_j
\approx E_{iso} \theta_j^2/2$.  One can derive the temporal evolution
of the peak frequency and peak flux during the lateral expansion
phase, by noting that during this phase, the jet opening angle scales
with the radius roughly as $\theta \propto R$, irrespective of the
initial jet opening angle, $\theta_j$ \citep{vEMWK11}. Conservation of
energy implies that
\beq
E_j \propto R^3 \theta^2 u_2 \propto R^5
\Gamma^2 \beta^2 = Const,
\label{eq:E1}
\eeq
where we used the fact that the energy density behind the shock front,
$u_2 \propto \Gamma^2 \beta^2$ (see Equation \ref{energy-density}),
which is valid for expansion into $r$-independent ISM density profile.

In the Newtonian regime, $\beta \ll 1$, which is a good approximation
during this phase, one thus finds that the temporal evolution of the
radius and velocity are $R \propto t^{2/7}$ and $\beta \propto
t^{-5/7}$.  Using these results in Equations \ref{magneticfield} and
\ref{electronenergy}, one finds $B \propto \beta \propto t^{-5/7}$ and
$\gamma_{\rm el} \propto \beta^2 \propto t^{-10/7}$.  Moreover, the
total number of radiating electrons collected from the ISM is $N_e
\propto R^3 \theta^2 \propto R^5 \propto t^{10/7}$.

Repeating the same arguments as in section \ref{sec:dynamics} above,
one finds that in the optically thin regime, $\nu_p^{\rm ob} \propto
B \gamma_{\rm el}^2 \Gamma \propto t^{-25/7}$ and $F_{\nu_p} \propto
N_e B \propto t^{5/7}$. In the optically thick regime, $\nu_p^{\rm
ob} \propto \nu_a^{\rm ob} \propto \Gamma B^{{P+2 \over P+4}}
\gamma_{\rm el}^{{2p-2 \over p+4}} R^{{2 \over p+4}} \propto
t^{14-25p \over 7(p+4)} \propto t^{-6/7}$, where the last equality
holds for $p=2$. Similarly, in this regime, $F_{\nu_p} \propto N_e
R^{-{p-1 \over p+4}} B^{2p+3 \over p+4} \gamma_{\rm el}^{5(p-1)
\over p+4} \Gamma  \propto t^{77-52p \over 7(p+4)} \propto
t^{-9/14}$ (for $p=2$).

\section{Interpretation of the late time radio emission of Sw J1644+57}
\label{sec:data}

\begin{figure}[ht!]
\centerline{\includegraphics[width=12cm]{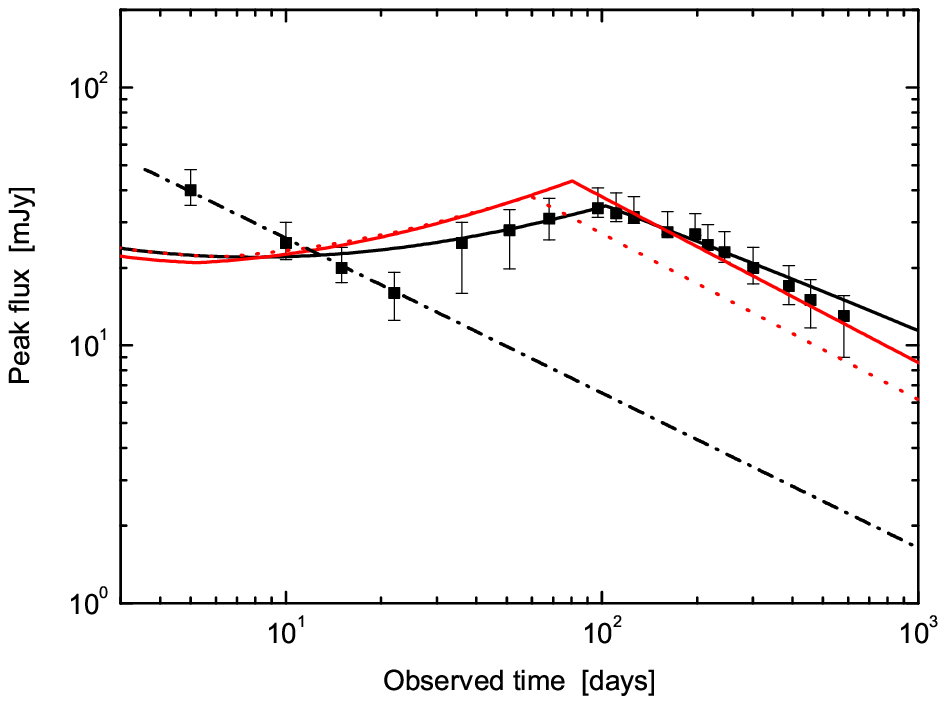}}
\caption{Temporal evolution of peak flux of the radio emission of
  Swift 1644+57. Data is taken from \citet{Berger+12} and
    \citet{Zauderer+13}. The vertical bars represent the uncertainties
    in determining the exact value of the peak flux from the data
    provided by \citet{Berger+12} and \citet{Zauderer+13}, and not
    statistical errors, which are not publicly available. Dash-dotted
  line is the contribution of the outer (slower) jet, while black
  solid line is the contribution from the inner (faster)
  jet in the spherical scenario.  The red lines represent fits
    to the data obtained when lateral expansion is considered, when
    the inner jet Lorentz factor reaches $\Gamma \leq 2$. The red
    solid line assumes the initial parameters described in \S4.3,
    while the red dotted line uses the same initial parameters as in
    the spherical scenario.  As the inner jet propagates into the
  ISM, it collects material and cools; thus, after $\sim 30$~days,
  radio emission from this region dominates the radio flux. At $\sim
  100$~days, the peak flux enters the optically thick regime
  ($\tau_{\nu_p} > 1$), which causes the decay seen at these
  times. The values of the free model parameters used are presented in
  the text.  }
 \label{fig:4}
\end{figure}

\begin{figure}[ht!]
\centerline{\includegraphics[width=12cm]{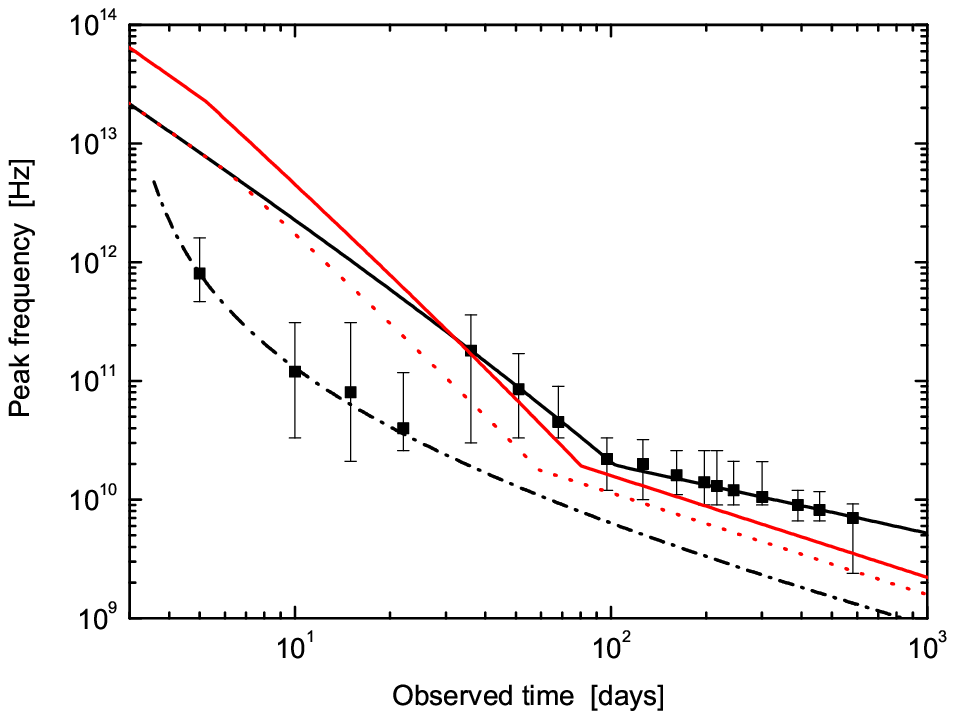}}
\caption{Temporal evolution of peak frequency for the radio emission
  of Swift 1644+57.  Observed data taken from \citet{Berger+12} and \citet{Zauderer+13}.
  Lines have the same meaning as in Figure \ref{fig:4}. Three
  distinctive regimes are clearly seen.  }
 \label{fig:5}
\end{figure}

The temporal evolution of the radio flux and peak radio frequency of
Sw J1644+57 up to 582 days from the initial outburst are presented
in Figures \ref{fig:4}, \ref{fig:5} \citep[data taken
  from][]{Berger+12, Zauderer+13}\footnote{The vertical bars in these figures
  represent the uncertainties in determining the exact values of the
  peak frequency and flux, which are derived from the data provided by
  \citet{Berger+12} and \citet{Zauderer+13}. No statistical errors on the values are publicly
  available.}. Three separate regimes are identified in both figures:
(I) At early times, $\lesssim 30$~days, the flux decreases from $\sim
40~{\rm mJy}$ at $\sim 5$~days to $\sim 15~{\rm mJy}$ at $\sim
30$~days. During this period, $\nu_p^{\rm ob}$ rapidly decays, with a
decay law consistent with $\nu_p^{\rm ob}\propto t^{-\alpha}$ and
$\alpha \approx 2.0$. (II) Between $30-100$~days, the flux {\it
  increases} by a factor of $\sim 1.8$ (from $\sim 20~{\rm mJy}$ to
$\sim 35~{\rm mJy}$). At the beginning of this epoch, at $t^{\rm ob}
\sim 30 - 45$~days, the radio peak frequency $\nu_p^{\rm ob}$
increases by a factor of a $few$, while during the rest of this epoch
it shows a similar decay law as is seen at early times, $\nu_p^{\rm
  ob} \propto t^{-\alpha}$ with $\alpha\approx 2.0$. (III) Finally, at
very late times, $\gtrsim 100$~days, the flux decays again. This
decay is accompanied by a slow decay in the peak radio frequency,
much slower than the decay observed at earlier phases. We point out
that the increase in the radio flux observed at epoch (II) led
\citet{Berger+12} to suggest that \textbf{late time energy
injection} may take place.

Based on the discussion presented in the previous sections, we
suggest the following interpretation to the late time radio emission
of Sw J1644+57: (I) During the early decay phase, $t^{\rm ob} < 30$
days, the emission is dominated by synchrotron radiation from the
same electrons that emitted the radio emission at early times; this
is the outer jet component in our model (see Figure \ref{fig:3}).
This emission is thus a continuation of the emission observed at
early times. The observed decay of both the flux and peak frequency
is due to the decreasing of magnetic field and the radiative cooling
of these electrons.  (II) At $t^{\rm ob} \approx 30$ days, there is
a transition: the inner jet plasma, that originally emitted the
X-ray photons, expands into the ISM, collects ISM material and
cools. At this time, synchrotron emission from this plasma becomes
the dominant component at radio frequencies. Initially, the inner
jet propagates at relativistic speeds, with $\Gamma \gtrsim 15$ (see
discussion in \S\ref{sec:X-rays} and Table \ref{tab01}). However, as
it propagates into the ISM, the plasma collects material from the
surrounding ISM and slows; the collected material contributes to the
radio emission, resulting in an increase in the radio flux.  (III)
Finally, at $t^{\rm
  ob} \sim 100$ days, the emission becomes optically thick (self
absorption frequency is larger than the peak frequency), which causes
the late time decay.

Fits to the temporal evolution of the peak radio frequency and peak
flux are shown in Figures \ref{fig:4} and \ref{fig:5}. The black solid
curves represent fits done using spherically symmetric scenario, while
in the red solid and dotted curves we fit the data using the lateral
expansion scaling laws derived in \S\ref{sec:lateral}, which are used
when the Lorentz factor drops below $\Gamma \leq 2$.

Consider first the spherical scenario, presented by the black
  solid curves.  In producing the fits, we use the following
parameters, which match the early and late time properties of the
flow. For the outer jet, we use initial expansion radius (at observed
time $t_0^{\rm ob} = 5$~days) $R_0 = 10^{16}$~cm. The typical
electron's Lorentz factor is taken to be $\gamma_{\rm el,0}=190$,
magnetic field $B_{0}= 4.5$~G, and initial bulk Lorentz factor
$\Gamma(t_0^{\rm ob}) \simeq 1.18$.  These values correspond to
$\epsilon_{\rm e} \approx 0.58$ and $\epsilon_{\rm B} \approx 0.13$,
and imply total number of radiating particles $N_{e} (t_0)=1.25\times
10^{53}$.  These values are consistent with the findings in
\S\ref{sec:radio1} (see Table \ref{tab01}). These parameters result in
initial peak synchrotron frequency $\nu_{p, 0}^{\rm ob}=8.0\times
10^{11}$~Hz, and peak observed flux $F_{\nu_{p},0}=39.4$~mJy.

For the inner jet, we find that the best fit is obtained when using
initial bulk Lorentz factor $\Gamma(t^{\rm ob} = 100~\rm{s}) =17$, $\epsilon_{\rm
  e} = 0.54$, $\epsilon_{\rm B} = 0.54$ and $r(t^{\rm ob} = 100~\rm{s}) \simeq
1.2 \times 10^{15}$~cm.\footnote{Note that $\epsilon_B$ represent the
  strength of the magnetic energy, normalized to the shock energy. As
  the magnetic field can have an external origin to the shock - e.g.,
  at the core of the disruptive star - it is possible that $\epsilon_e
  + \epsilon_B > 1$.}
\footnote{We took initial parameter values derived at $t^{\rm ob} =
  100~\rm{s}$, to be consistent with the analysis in section 2 above.}
 We point out that the value of $\epsilon_B$ is
not constrained by early X-ray data.  Moreover, here the collected ISM
plays a significant role. Thus, we could constrain the ISM density to
be $n_{\rm ISM}=2.7~{\rm cm^{-3}}$. We further assume that the shocked
ISM have a power law distribution above $\gamma_{\rm el}$ with power
law index $p=2$. When calculating the dynamics in equation
(\ref{dynamicalevolution}), we consider adiabatic expansion, namely
$\epsilon=0$.

\subsection{The transition at 30 days}

Within the context of our model there are four separated
  regimes that contribute to the emission.  (I) Emission from matter
  ejected at the outer (slower) jet dominates before $\sim 30$ days,
  as is shown by the fits; (II) contribution from ISM material
  collected by the outer jet, which is sub-dominant, as the amount of
  ISM material collected in the first month is significantly smaller
  than the material that originated in the outer jet (see discussion
  in \S\ref{sec:cooling}); (III) contribution of emission from
  material originated in the inner jet (which, as we show below, can not be constrained
  by the data); and (IV) contribution of ISM material collected by the
  inner jet, which becomes the dominant source of radio emission after
  $\sim 30$ days, and causes the rise in the radio flux observed at
  later times.

At $\approx 30$ days a transition occurs, as contribution from the
inner jet (and ISM material collected by it) becomes the dominant
source of radio emission. Using equations \ref{dynamicalevolution},
\ref{eq:cm}, one finds that at that time, the inner jet is at radius
$\sim 4.2\times 10^{17}$~cm and the number of the collected ISM
protons is $\sim 8.4\times 10^{53}$, which is much larger than the
number of particles that initially existed in the jet, $N_{\rm
  initial}=5.5\times 10^{52}$ in the fits presented. Therefore, the
contribution of the collected ISM to radio emission from the inner
jet is dominant (under the assumption of similar magnetic fields
  and similar Lorentz factor of the radiative electrons at both the
  ejected plasma and the collected ISM).

 We note that a significant deceleration of the inner jet
   occurs much earlier. The deceleration becomes significant when the
   collected mass from the ISM is $\approx N_{\rm
     initial}m_{p}/\Gamma$ \citep[e.g.,][]{Peer12}. This occurs at
   radius $R\sim 7\times 10^{16}$~cm, corresponding to observed time
   $R/(\Gamma^{2}c) \approx 8000$~s (assuming initial inner jet
   properties $\Gamma =17$ and $N_{\rm initial}=5.5\times 10^{52}$, as
   discussed above and in Figure \ref{fig:2}). Therefore, a
   significant decrease in the inner jet Lorentz factor occurs already
   after a few hours. Using Equations \ref{dynamicalevolution},
   \ref{eq:cm}, we find that the inner jet Lorentz factor drops to
   $\Gamma \leq 2$ after $\approx 6$~days (observed time). At this
   epoch, lateral expansion may be significant; see discussion in
   \S4.3 below.

\begin{figure}[ht!]
\centerline{\includegraphics[width=12cm]{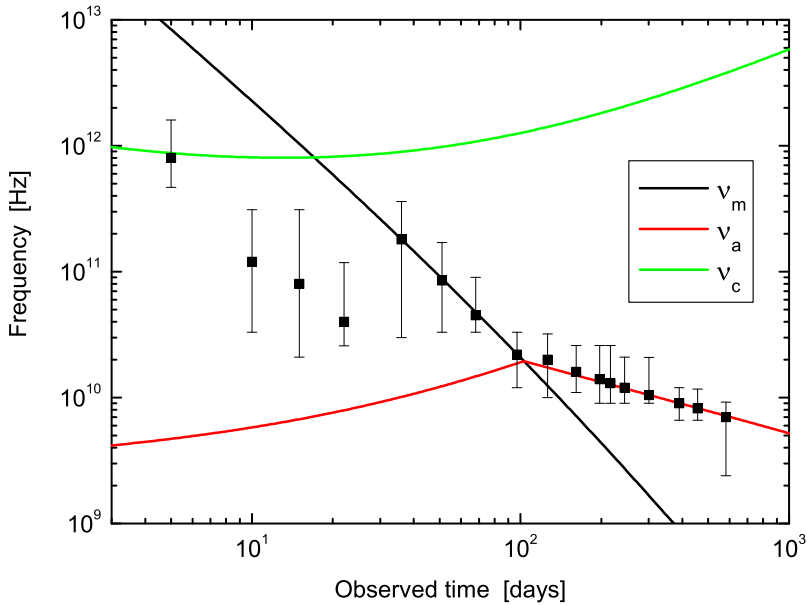}}
\caption{Temporal evolution of characteristic synchrotron frequencies:
  peak frequency $\nu_m$, self absorption frequency $\nu_a$ and
  cooling frequency, $\nu_c$ of emission from the inner jet, are
  plotted on top of the data. At the epoch 30-100 days, the emission
  is in the optically thin region, while at $t^{\rm ob}>100$~days it
  is in the optically thick regime. In producing the figure,
    we used the same parameters as in the spherical scenario, namely
    $\Gamma_{\rm initial} =17$, $N_{\rm initial}=5.5\times 10^{52}$
    and $n_{\rm ISM}=2.7~{\rm cm}^{-3}$.}
 \label{fig:6}
\end{figure}

The collected ISM contributes to the radio emission from the inner
jet. Its contribution to the radio flux becomes comparable to that
of the outer jet after $\gtrsim 10$ days, and dominant after $\sim
30$ days (see Figures \ref{fig:4}, \ref{fig:5}). At early times,
however, the radio emission from the inner jet peaks at higher
frequencies than those observed (see Figures \ref{fig:5} and
\ref{fig:6}). Figure \ref{fig:6} shows the temporal evolution of the
three
  characteristic synchrotron frequencies: peak frequency $\nu_m$, self
  absorption frequency $\nu_a$ and cooling frequency, $\nu_c$ of
  emission from the inner jet. In the figure we use the same
  parameters as of the spherical scenario discussed above, i.e.,
  $\Gamma_{\rm initial} =17$, $N_{\rm initial}=5.5\times 10^{52}$ and
  $n_{\rm ISM}=2.7~{\rm cm}^{-3}$.  For example, at $\sim 20$~days,
emission from the inner jet is expected to be peaked at $\sim
6\times 10^{11}$~Hz, with peak flux of $F_{\nu_p} \approx 20$~mJy.
Available data at these times is limited to lower frequencies, $\leq
230$~GHz \citep{Berger+12}. The observed flux at 230 GHz, $\approx
10$~mJy, is lower than the expected flux at the peak frequency of
$\sim 600$~GHz, as it should; It is in agreement with the
expectations from synchrotron theory, $F_\nu/F_{\nu_p} =
(\nu/\nu_p)^{1/3}$ for $\nu_a<\nu<\nu_p$. Since at an even earlier
times the peak flux of radio emission from the inner jet is at
higher frequencies (see Figure \ref{fig:6}), its contribution at the
flux at the observed bands of 230 GHZ and below is further reduced.

 At $\sim 30$ days, the peak frequency from the inner jet (And the ISM
 material collected by it) is $\nu_m^{\rm ob}\approx 2.5\times
 10^{11}$~Hz and the self-absorption frequency is $ \nu_{a}\approx
 1.0\times 10^{10}$~Hz, namely the radio emission is in the {\it
   optically thin} regime (see Figure \ref{fig:6}).
We point out that
 these results are, in fact not far from the fits done by
 \citet{Berger+12} (see their table 2) for that time epoch, and
 therefore our model can equally well fit the existing spectrum
 \citep[Figure 2 in][]{Berger+12}.

\subsection{The transition at 100 days}

The peak emission frequency (From the inner jet) drops faster
  than the self absorption frequency (see discussion in section 3.1
  above, and Figure \ref{fig:6}).  At observed time $t^{\rm ob} = 100$~days, the radio
  emission becomes {\it optically thick}, implying a change in the
  temporal decay of the observed peak flux and peak frequency, which
  are consistent with observations.

Using Equations \ref{dynamicalevolution}, \ref{eq:cm}, we find that
at observed time $t^{\rm ob} = 100$~days, the inner jet reaches
radius $R\sim 6.0\times 10^{17}~{\rm cm}$, with bulk Lorentz factor
$\Gamma\sim 1.15$. At this time, the magnetic field is $B\sim
0.19$~G and the characteristic electrons Lorentz factor is
$\gamma_{\rm el} \approx 150$. These result in $\nu_p^{\rm ob} =
2.06\times 10^{10}$~Hz and $F_{\nu_p} = 34.4$ mJy. However, at this
time, $\tau_{\nu_{p}} \approx 1$, which implies that from this time
onward synchrotron radiation is in the optically thick regime, and
its late time properties are described by equations
(\ref{peakfrequency2}). The black solid lines in Figures
\ref{fig:4}, \ref{fig:5} represent the expected peak frequency and
flux in the optically thin region at early times, and optically
thick regions at $t^{\rm ob} > 100$~days.

It should be emphasized that the fits presented here are
  different than those used by \citet{Berger+12} and
  \citet{Zauderer+13}. In these works, spectral fits for the long-term
  radio data (up to $\sim$ 582 days) was done based on the formulae of
  the synchrotron spectrum model described in \citet{MGM12}, or
  equation 4 in \citet{Berger+12}. These, in turn, assume relativistic
  motion, $\Gamma \gg 1$.  The obtained values of the fitted
  parameters vary with time; variation of the luminosity had led
  \citet{Berger+12} to propose late time energy injection.
  Furthermore, \citet{Berger+12} found $\nu_{a}<\nu_{m}$ even at late
  times. However, we point out that in their fits, they find $\nu_{a}
  \approx \nu_{m}$ at these epoch (see their table 2). Since the
  spectra at high frequencies $\nu \gg \max\{\nu_a, \nu_m\}$ and at low
  frequencies, $\nu \ll \min\{\nu_a, \nu_m\}$ is similar in the
  optically thin and optically thick cases, it is of no surprise that
  our model is also capable of reproducing the spectra.

In our model the late time radio emission after $\sim 30$~days
originates from the interaction between the inner jet with ISM.
Using the initial mass and bulk Lorentz factor of inner jetted
outflow obtained by the analysis for the early X-rays in
\S\ref{sec:early_times}, we obtain good fits to the late time radio
data presented in Figures \ref{fig:4}, \ref{fig:5}, without the need
for late energy injection.

\subsection{Lateral Expansion}

We next consider a scenario in which lateral expansion takes place.
These are shown by the red solid and dotted curves in Figures
\ref{fig:4}, \ref{fig:5}. Such an expansion will take place once a
jetted outflow slows down to $\Gamma \sim 1/\theta_j$. Since the
initial jet opening angle $\theta_j$ is not constrained by the data,
we cannot evaluate the significance of this effect. We
  therefore rely on the results of the recent simulations carried by
  \citep{vEvdHM12}, which indicate that sideways expansion becomes
  significant for $\Gamma \simeq 2.0$. Thus, in our fittings, the
  lateral expansion is considered when the Lorentz factor of the inner
  jet drops below 2. As explained above, this is expected after
  $\approx 6$~days (observed time).


When fitting the data, we use the scaling laws described in
\S\ref{sec:lateral} to calculate the fits for temporal evolution of
the peak radio frequency and peak flux after $\sim 30$ days. We make
two fits to the data, shown by the red solid and red dotted lines in
Figures \ref{fig:4} and \ref{fig:5}. In the solid curve, we use the
physical parameters $\Gamma(t_0^{\rm ob}) =15$, $\epsilon_{\rm e} =
0.84$, $\epsilon_{\rm B} = 1.65$, $n_{\rm ISM}=2.6~{\rm cm}^{-3}$ and
$r(t_0^{\rm ob}) \simeq 1.0 \times 10^{15}$~cm. We find that at
observed time $t^{\rm ob} \approx 81$~days, the radio emission enters
the optically thick regime with peak frequency $\nu_p^{\rm ob} =
1.92\times 10^{10}$~Hz and peak flux $F_{\nu_p} = 43.4$ mJy.  At this
time, the outflow reaches radius $R\sim 3.63\times 10^{17}~{\rm cm}$,
with bulk Lorentz factor $\Gamma\sim 1.074$, magnetic field $B\sim
0.25$~G and characteristic electrons Lorentz factor $\gamma_{\rm el}
\approx 127$.

In order to estimate the sole effect of the lateral expansion,
  we show in the red dotted curve the results of the fit made using
  similar parameters to those adopted in the spherical scenario, with
  lateral expansion taking place for $\Gamma \leq 2$.  The results of
  the fits indicate that lateral expansion has a minor effect on the
  obtained data.


We can thus conclude that the results presented in Figures
\ref{fig:4} and \ref{fig:5} indicate that both the spherical
expansion and the sideways expansion scenarios enable us to obtain
good fits to the late times radio data, within the context of the
two component jet model proposed here.  Moreover, we find that in
both scenarios, similar values of the free model parameters are
obtained.


\section{Conclusions and Discussions}
\label{sec:summary}

In this paper, we studied the early and late time emission of the
TDE Sw J1644+57, both at radio and X-ray frequencies. Based on our
analysis, we propose a two-component jet model (see Figure
\ref{fig:3}) that fits the observations. Our model contains
  both dynamical and radiative parts, and give a satisfactory
  interpretation for the origin of {\it both} the early time X-ray and
  radio emission, as well as the complex late time radio
behavior. This model also predicts that the GeV emission, originated
from second Compton scattering of the X-ray photons in the inner
jet, is markedly suppressed by $\gamma -\gamma$ annihilation. While
  we present here the basic ingredients of the model, as well as the key
  break frequencies, we leave a detailed study of the full temporal
  and spectral evolution to a future work.

By analyzing the early time ($t^{\rm ob} \lesssim 5$~days) data, we
conclude that the radio and X-ray photons must have separate origins
(see \S\ref{sec:early_times}, Table \ref{tab01}). This conclusion is
based on the assumption that the X-rays are produced by relativistic
electrons through inverse Compton scattering of the radio photons.
We are able to put strong constraints on the properties of the radio
emitting plasma (see \S\ref{sec:radio1} and Figure \ref{fig:1}), and
somewhat weaker constraints on the properties of the X-ray emitting
plasma (\S\ref{sec:X-rays} and Figure \ref{fig:2}). Stronger
constraints on the {\it initial} properties of the X-ray emitting
plasma are obtained when considering late time radio emission.

The results of our analysis, as presented in Table \ref{tab01},
indicate that the electrons in the outer jet region are hotter than
those in the inner jet, namely a larger fraction of kinetic energy
is used to accelerate electrons to relativistic energies in the
outer jet region. This is in spite of the fact that the outer jet
region is slower than that of the inner jet. However, we point out
that it is possible that the slower, outer jet is simply the
boundary layer between the fast jet and the ambient medium. Since
viscous friction between the fast jet core and the static ambient
medium converts kinetic energy into heat, it would be natural for
the slow outer jet to be hotter.  Moreover, understanding of
particle acceleration in (trans)-relativistic outflows is far from
complete, and currently no theory is fully developed that it enables
a connection between the bulk outflow motion and the characteristic
energy of the accelerated electrons. Our results, based on data
analysis, could thus serve as a guideline in constructing such a
theory.

Our jet within a jet model naturally explains the complex temporal
evolution of the radio emission at late times (up to $\sim
600$~days). We solved in \S\ref{sec:late_times} the dynamics and
expected radiation from the jet propagating into the ISM. We stress
that as the initial Lorentz factor of the inner jet is mild (our
best fit gives $\Gamma \approx 17$), one has to consider the
transition between the relativistic and the Newtonian expansion
phases, and cannot rely on analysis of ultra-relativistic outflows,
as is the case in GRBs. We further consider the radiative cooling of
particles behind the shock front.

We demonstrated how our model can be used to fit the data in
\S\ref{sec:data}. Our key idea is that the observed signal can be
split into three separate regions: at early times, radio emission is
dominated by the outer jet. The decay of the peak frequency and flux
is attributed to radiative cooling of the electrons and the
declining of magnetic field.  Between $30-100$~days, the radio
emission is dominated by the inner jet, in the optically thin
regime. The inner jet propagates at relativistic speeds and collects
material from the surrounding ISM.  The addition of the collected
material results in the increase of radio flux. Finally, at very
late time, $t^{\rm ob} \gtrsim 100$~days, the radio emission is
dominated by the inner jet, in the {\it optically thick} regime.
This causes the observed late time decay of peak flux and frequency.
Our conclusions are not changed when considering lateral expansion
of the inner jet. Comparing the results of spherical expansion and
lateral expansion, we find that the values of the free model
parameters are similar. We point out that in the late stage ($t^{\rm
ob} \gtrsim 100$~days) of a lateral expansion, the expansion becomes
spherical \citep{vEvdHM12}.

When analyzing the early time data, we use the common interpretation
of the radio emission as originating from synchrotron radiation. We
find that the bulk Lorentz factor of the synchrotron emitting plasma
is $\Gamma\lesssim 1.2$. Our analysis method is similar to that used
by \citet{Zauderer+11}, albeit being more general, as we avoid the
equipartition assumption used in that work. The X-ray emission is
interpreted as IC scattering of the radio photons, however our
analysis indicates that the bulk motion of the X-ray emitting region
is much higher, $\Gamma \gtrsim 15$; alternatively, the emission
radius is much smaller, $r \sim 10^{12.5}$~cm.  We thus conclude that
radio and X-ray emission zones are completely separated, either
spatially or in velocity space. This is consistent with the separate
temporal behavior observed at early times, as the observed X-rays
maintained a more constant level after the first 48 hours
\citep[albeit with episodic brightening and fading spanning more than
  an order of magnitude in flux; see ][]{Levan+11}, whereas the low
frequency emission decreased markedly.

Within the context of our model, one can estimate the total energy
of both the inner and outer jets at early times, using the fitted
  parameters obtained in \S\ref{sec:early_times}.  The (initial)
kinetic energy of the inner jet is $E_{\rm initial} = N_{\rm
  initial}(\Gamma m_{p}c^{2}) \approx 3\times 10^{51}$~erg, where we
have used $\Gamma\sim 20$ and $N_{\rm initial}\sim 10^{53}$ (see
Table 1).  This value is derived by considering material ejection
  during an observed variability time of $\sim 100$~s. As the averaged
  observed X-ray luminosity during the first three days is $\sim
  3\times 10^{47}~{\rm erg~s^{-1}}$, with flaring activities (lasting
  $\lesssim 100$~s) of about an order of magnitude higher
  \citep{Burrows+11}, we conclude that the kinetic energy in the inner
  jet is at least an order of magnitude larger than the observed
  energy in an X-ray flare. There are several dozens of flaring
  activities observed at the X-ray band during the first three days, during which
  total energy of $\gtrsim 10^{53}$~erg is inferred from observations at this band
  \citep{Burrows+11}. Thus, conservatively, we can conclude that the
  (isotropic equivalent) kinetic energy injected in the inner jet
  during this period is $> 10^{54}$~erg.

For the outer jet, the estimated kinetic energy is $(\Gamma
  -1)m_{p}c^{2}N_{e} \approx 3.4\times 10^{49}$~erg, where $\Gamma
  =1.18$ and $N_{e}=1.25\times 10^{53}$ are taken, and the electron's
  thermal energy is $\gamma m_{e}c^{2}N_{e}=1.95\times 10^{49}$~erg
  (using $\gamma_{e}=190$). This gives outer jet energy of $\approx
  5.4\times 10^{49}$~erg.  The averaged observed luminosity of the
  radio emission within the first 5 days is $ \approx 3\times
  10^{43}~{\rm erg~s^{-1}}$, implying total energy release at radio
  frequencies of $\sim 10^{48.5}$~erg during this period. We can
  therefore conclude that the estimated kinetic energy contained in
  both the inner as well as the outer jets in our model is at least an
  order of magnitude larger than the observed energy released at radio
  and X-bands, more than enough to explain these
  observations. Moreover, we point out that these values are
  consistent with a few - few tens of percents efficiency in
  conversion of kinetic energy to radiative energy estimated in
  observations of jets from gamma-ray burst \citep[e.g.,][]{Peer+12}.

Our calculated X-ray luminosity can match the observed X-ray
luminosity of TDE Sw J1644+57, $L_{X}\approx 10^{47.5}~{\rm
  erg~sec^{-1}}$, exceeding the radio emission by a factor $\sim
10^{3}-10^{4}$. If the emission is isotropic, the X-ray luminosity of
Sw J1644+57 corresponds to the Eddington luminosity of an $\sim
10^{9}~M_{\odot}$~MBH, which is incompatible with the upper limit
$\sim 10^{7}~M_{\odot}$ of the MBH mass derived from variability
\citep{Bloom+11, Burrows+11}, so the source is required to be
relativistically beamed. We have obtained the bulk Lorentz factor of
the relativistic jetted outburst $\Gamma\sim 15$, which is very close
to the typical value inferred in blazars \citep{Jiang+98}.  Therefore,
if the beaming angle of the jet $\theta_{j}\approx 1/\Gamma\sim 0.1$,
the beaming-corrected luminosity $f_{\rm b}L_{X}\sim 10^{45}~{\rm
  erg~sec^{-1}}$ ($f_{\rm b}=\left (1-\cos\theta_{j}\right )$ is the
beaming factor) becomes consistent with the Eddington luminosity of a
$\sim 10^{7}~M_{\odot}$~MBH \citep[see also ][]{Bloom+11}.

\citet{Berger+12} and \citet{Zauderer+13} present the continued
radio observations of the TDE Sw J1644+57 extending to $\sim 582$
days. In the work of \citet{Berger+12}, they fitted the data using
the model of \citet{GS02}, and concluded that the re-brightening
seen after $\sim 30$ days cannot be explained in the framework of
that model. They thus conclude that an increase in the energy by
about an order of magnitude is required. As discussed above, the
re-brightening is very natural in our scenario due to the increase
in the collected material when the jetted outburst propagates
through the ISM.

Recent observation \citep{Zauderer+13} reveals a sharp drop in the
X-rays flux of this object at time $\delta t\lesssim 500$ days, while
a similar behavior in the radio emission is absent. This result
supports our conclusions that the radio and X-ray emission have
separate origins (see \S\ref{sec:early_times}, Table \ref{tab01}). In
our model, the early X-rays originate from the IC scattering of radio
photons by fast electrons in the inner jet, possibly located at the
radius $r\sim 10^{15}-10^{16}~{\rm cm}$ (see Table \ref{tab01}),
consistent with the result of \citet{Decolle+12}. These electrons
later cool, contributing to the radio emission.

In the framework of the two-component jet model, the SSC process in
outer jet cannot be responsible for the early X-ray emission, as
discussed in \citet{Zauderer+11}. We have computed the SSC emission
from electrons that produce the radio emission at $\sim 500$ days
(the outflow reaches the radius $\sim 10^{18}$ cm), and find the
peak occurs $\sim 10^{-3}-10^{-2}$~eV and the luminosity at the peak
is about $\sim 2.0\times 10^{38}~{\rm erg~s^{-1}}$, which are both
much lower than the observed values.

In the framework of our model, we thus consider two possible
explanations of the rapid decline in the X-ray emission.  One
possibility is that the late time X-ray emission has a completely
separate origin, e.g., from the accreted stellar debris. In such a
scenario, shut off of the inner engine can lead to the decay of the
X-ray flux; this idea is somewhat similar to that discussed by
\citet{Zauderer+13}.  Alternatively, if the origin of the late-time
X-rays is SSC by electrons accelerated in the inner jet, possibly
located at the disrupted radius $r\sim 10^{15}-10^{16}~{\rm cm}$
(note that late time radio and X-ray emission have different
emission regions. At $\sim 500$ days, the jetted material for
producing late radio emission reaches the radius $\sim 10^{18}$ cm),
the rapid decline can potentially be explained by jet precession,
which forces the inner jet to move away from our sight.

The idea of jet within a jet was suggested in the past as a way to
explain the morphology of high energy emission from active galactic
nuclei (AGNs) \citep{GTC05, Hardcastle06, Jester+06, Siem+07}, and in
the context of GRBs in explaining the break observed in the afterglow
light curve \citep{Racusin+08, DePasquale+09, Filagis+11}, as well as
some of the properties of the prompt emission \citep{LPR13, LPR14}.
While the theory of jet launching is still incomplete, clearly, jets,
being collimated outflows are expected to have lateral velocity
gradient. The analysis done here suggests that such a velocity
gradient exists in jets originating from TDEs. In principle, the outer
jet may also represent a cylindrical boundary layer owing to the
interaction of the inner jet with the ambient gas.

Recently, a two component jet model for Sw J1644+57 was
  proposed by \citet{Wang14}. In difference from our model,
  the early radio emission is assumed to originate from the inner jet,
  while the outer jet is responsible for the late radio emission; both
  the inner and outer jets contribute to the X-ray emission.

While Sw~J1644+57 is the first TDE event from which the existence of
relativistic jets is inferred, and the most widely discussed one in
this context, it is possible that other such events were
detected. Recently, \citet{Cenko+12} reported a second event,
Sw~J2058.4+0516 which is a potential candidate for a relativistic
flare. While currently, no late time radio lightcurve is currently
available, we can predict that if such a lightcurve becomes available
it should show the same complex behavior of Sw J1644+57.

\acknowledgments We would like to thank the anonymous referee for many
useful comments that helped us improve this manuscript.  We are
grateful to Bevin A. Zauderer and Brian D.  Metzger for useful
discussions. DL acknowledges support by the National Science
Foundation of China (Grant Nos. 11078014 and 11125313), the National
Basic Research Program of China (Grant Nos.  2009CB824904 and
2013CB837901), and the Shanghai Science and Technology Commission
(Program of Shanghai Subject Chief Scientist; Grant Nos. 12XD1406200
and 11DZ2260700). This work was supported in part by NSF grant
AST-0907890 and NASA grants NNX08AL43G and NNA09DB30A. AP acknowledges
support from Fermi GI program \#41162.  This work was supported in
part by National Science Foundation Grant No. PHYS-1066293 and the
hospitality of the Aspen Center for Physics.







\end{document}